# Does ridesourcing respond to unplanned rail disruptions? A natural experiment analysis of mobility resilience and disparity

**Elisa Borowski, Jason Soria, Joseph Schofer, Amanda Stathopoulos\***
Department of Civil and Environmental Engineering
Northwestern University, Technological Institute, 2145 Sheridan Road, Evanston, IL 60208 USA
\* a-stathopoulos@northwestern.edu

**Abstract**
 Urban rail transit networks provide critical access to opportunities and livelihood in many urban systems. Ensuring that these services are resilient (that is, exhibiting efficient response to and recovery from disruptions) is a key economic and social priority. Increasingly, the ability of urban rail systems to cope with disruptions is a function of a complex patchwork of mobility options, wherein alternative modes can complement and fill service gaps. This study analyzes the role of ridesourcing in providing adaptive mobility capacity that could be leveraged to fill no-notice gaps in rail transit services, addressing the question of distributional impacts of resilience. Using a natural experiment, we systematically identify 28 major transit disruptions over the period of one year in Chicago and match them, both temporally and spatially, with ridesourcing trip data. Using multilevel mixed modeling, we quantify variation in the adaptive use of on-demand mobility across the racially and economically diverse city of Chicago. Our findings show that the gap-filling potential of adaptive ridesourcing during rail transit disruptions is significantly influenced by station-, community-, and district-level factors. Specifically, greater shifts to ridesourcing occur during weekdays, nonholidays, and more severe disruptions, in community areas that have higher percentages of White residents and transit commuters, and in the more affluent North district of the city. These findings suggest that while ridesourcing appears to provide adaptive capacity during rail disruptions, its benefits do not appear to be equitable for lower-income communities of color that already experience limited mobility options. Research implications for mobility operator collaborations to support mobility as a service are discussed. This study builds a more comprehensive understanding of transit service resilience, variation in vulnerability, and the complementarity of ridesourcing to existing transport networks during disruptions.

**Keywords:** mobility resilience; transit disruption; multilevel model; natural experiment; ridesourcing; community equity

**1. Introduction**

Urban livability refers to the quality of life in urban communities and the degree to which cities satisfy the needs and aspirations of their inhabitants by providing physical and social well-being and supporting meaningful existences (Kashef, 2016). One important component of urban livability is a resilient mobility system that provides reliable access to work, healthcare, food, recreation, and other life-sustaining services. Ensuring resilience is challenging, however, because transportation systems rely on a complex web of fixed assets and multiple dynamic components, including competing operators, fixed-schedule and on-demand modes, and operations across heterogeneous built and social environments. In some regions, rail transit serves as a backbone for urban mobility (Litman, 2007). As such, in complex urban mobility systems, the ability of alternative modes to fill no-notice gaps in transit services is of critical importance, and the interplay between rail transit with other modes in the transportation system is increasingly recognized as foundational to mobility resilience.

Unexpected transit disruptions, such as service interruptions due to accidents, infrastructure breakdowns, and passenger distress, are common occurrences in urban transit systems. The presence of ridesourcing services in mobility portfolios introduces a novel opportunity to bolster mobility response and recovery. Recent work has begun to highlight the need for more equitable resilience plans, noting that impacts can vary across population groups and be tied to vulnerability (Coleman et al., 2020; Dargin & Mostafavi, 2020). In this research, we consider the role of ridesourcing as an adaptive substitution strategy to fill gaps created by no-notice disruptions in fixed guideway systems, such as rail transit services. Moreover, we analyze socio-spatial variation in rail-to-ridesourcing substitution across communities. Therefore, our discussion of resilience focuses on the dimension of redundancy, as well as distributional impacts of resilience. This is in line with other works that have identified the feature of redundancy across spatially heterogenous urban transport systems to reduce the risk of service disruptions (Estevez-Mauriz et al., 2017).

Our research uses a natural experiment to study the resilience of urban transit in Chicago. The natural experiment approach unlocks the opportunity to study variation that naturally results from an intervention beyond researcher control (e.g., Lu et al., 2021; Wen et al., 2017). By jointly studying naturally occurring disruptions in the rail transit system with associated surges in ridesourcing demand, we empirically capture the substitutability among competing modes. We systematically identify 28 significant no-notice rail transit disruptions over the period of one year (November 2018 through October 2019). Then each event is temporally and spatially matched with ridesourcing trip data from the City of Chicago. We estimate a multilevel mixed (MLM) model to examine socio-spatial variation in ridesourcing demand surges that are associated with transit disruptions across stations, neighborhoods, and districts, while controlling for the time of day, day of the week, and disruption location. The research design thereby captures spontaneous mobility resilience as travelers use ridesourcing as an adaptation strategy to cope with disruptions.

The main contributions of this study are the insights it provides into: (1) whether ridesourcing is used as a gap-filling mechanism during unplanned rail transit disruptions in Chicago, (2) whether its utilization for this purpose is equitably distributed across the city, especially in terms of racial and economic representation, and (3) whether variation in adaptive ridesourcing demand during disruptions is primarily attributable to station-, community-, or broader district-level factors. This study contributes to an improved understanding of how riders cope with disruptions across different communities. Furthermore, we discuss how our findings can guide more equitable communication strategies for transportation agencies and potential collaborations between private, on-demand mobility service operators to treat mobility as a service regardless of the specific transportation mode used.

Our research contributes to filling two main gaps in knowledge. First, considering the resilience concept of redundancy, we analyze how the complex multimodality of transportation systems contributes to mobility resilience in terms of the *interaction* between public transit and ridesourcing, rather than focusing on single-mode resilience. Second, acknowledging the growing interest in equitable resilience, we provide further insights on the socio-spatial variation in adaptive ridesourcing during unplanned transit disruptions. Specifically, to identify potential resilience inequities, we examine the variation of resilient mobility behavior across communities at three levels of aggregation. Our analysis suggests that under-resourced transit riders may not be benefitting as much as more privileged transit riders from the same ridesourcing-based, mobility gap-filling strategy.

The remainder of this paper is organized as follows. The next section reviews related studies on the relationships between transit disruptions, connections between rail transit and ridesourcing, and social equity. The third section discusses the data sources, variable definitions, and model specification. The fourth section compares the results from the MLM models and discusses the empirical findings. Research implications are discussed in the fifth section. Finally, the conclusion is presented in the sixth section.

## 2. Literature review

### 2.1. Mobility resilience

To study mobility resilience, we must consider how transit agencies are preparing for, reacting to, and recovering from service disruptions which impact the daily lives of individuals and businesses. Resilience is currently a key priority in national policy agendas and discourse (National Academies of Sciences, Engineering, and Medicine. (2021a). Mobility resilience refers to the adaptive ability of a transportation system to maintain functionality despite a disturbance (Walker et al., 2002; Walker et al., 2004). Mata and Martins (2019) define *mobility resilience* as the ability of travelers to maintain their mobility without compromising quality of life, *adaptable resilience* as the possibility of adopting new mobility patterns to maintain quality of life, and *transformable resilience* as the ability to transform current mobility patterns into more adaptable ones. Current research on mobility resilience covers a broad range of subjects, from post-disaster recovery following extreme climate events (Aydin et al., 2017; Chan & Schofer, 2016; Donovan & Work, 2017) to disruption preparation that takes into consideration network redundancies and impacts to links and nodes (Capri et al., 2016; Fotouhi et al., 2017; King & Shalaby, 2016; Leu et al., 2010; Serulle et al., 2011). Operationally, transportation resilience is difficult to define and measure (National Academies of Sciences, Engineering, and Medicine, 2021b), and many metrics have been proposed, often centered on the four Rs of resilience: *robustness* (i.e., maintaining operations while withstanding stresses), *resourcefulness* (e.g., supply chain management, communication, and mobilization of resources), *rapidity* (i.e., quickly returning to normal operations, containing losses, and preventing further degradation of the system), and *redundancy* (e.g., providing backup resources and substitutability) (Adams et al., 2012; DiPietro et al., 2014; Faturechi & Miller-Hooks, 2015; Gu et al., 2020). In this study, we focus on disruption recovery by applying the resilience dimension of *redundancy*, that is, the provision of backup urban mobility resources for modal substitution. Recently, resilience equity and its ties to vulnerability has attracted attention (Coleman et al., 2020; Dargin & Mostafavi, 2020). From a social equity perspective, it is important to consider how spatial and temporal differences in resilience can result in inequitable outcomes (Meerow & Newell, 2019). In this work, distributional patterns of modal substitution are analyzed to pinpoint these variations.

### 2.2. Recovering from mobility disruptions: What do we know?

Understanding mode substitution for disruption recovery requires us to ground our analysis on travel demand changes and an assessment of the known connections between disruptions in fixed transit services and on-demand ridesourcing behaviors.

#### 2.2.1. Recovery from planned rail transit disruptions: Decisions of travelers

The duration of planned transit service disruptions can span from short-term (e.g., hours or days) to long-term (e.g., months or years) (Arslan Asim et al., 2021). Planned, long-term rail transit disruptions may include scheduled maintenance, infrastructure upgrades, and strikes. For planned and long-term rail line or station closures, riders have ample time to adjust their travel plans, whether temporarily or permanently, including choice of departure time, route, and mode. The impacts of long-term rail transit disruptions on travel behavior have been widely studied over past decades (Marsden & Docherty, 2013; Pnevmatikou et al., 2015; Pu et al., 2017; van Exel & Rietveld, 2001; Zhu et al., 2017). Across Europe and the United States, effects of long-term transit disruptions have resulted in a permanent decline in transit ridership (van Exel & Rietveld, 2001; Zhu et al., 2017). In Chicago, lengthy track operation disruptions have led to transit abandonment by an estimated 4% of riders (Mojica, 2008).

In general, mode-shifting behavior during long-term transit disruptions has been shown to vary according to rider sociodemographics, disruption type, and city-specific factors. More specifically, transit disruption responses during strikes include individual-specific factors (i.e., car ownership, number of household members, available cars, driver's licenses, and income), context-specific factors (i.e., travel distance, travel time, travel cost, trip destination, and weather), and journey-specific factors (i.e., public transport station accessibility and trip purpose) (Nguyen-Phuoc et al., 2018a). Typically, during long-term disruptions, most travelers switch to personal vehicles (van Exel & Rietveld, 2001; Zhu et al., 2017), but disruption outcomes are often inequitable. For example, those less likely to shift to car during transit disruptions include women and lower-income individuals, as well as workers with more flexible schedules (Pnevmatikou et al., 2015). In Washington, D.C., long-term rail transit disruptions are associated with increased bus ridership (Pu et al., 2017), especially among lower-income riders (Zhu et al., 2017). In Chicago, planned maintenance resulted in a minor share of rail riders shifting to bus and the majority continuing to use rail transit (Mojica, 2008).

*2.2.2. Recovery from unplanned, short-term rail transit disruptions*

Compared to the extensive body of research on planned, long-term rail transit disruptions, research on the travel behavior effects of unplanned, short-term disruptions is scant (Sun et al., 2016). Unplanned service disruptions may be the result of extreme weather or natural disasters, infrastructure failures (e.g., related to power, signaling, and crossovers), vehicle breakdown (e.g., rolling stock issues), and service interruptions or delays due to intrusions on rail tracks or medical emergencies (e.g., debris, suicides, crossing incidents, etc.) (Arslan Asim et al., 2021; Pender et al., 2013). Important factors influencing behavioral responses to no-notice disruptions include user expertise, car availability, perception of service recovery time, available transport services, time constraints, and the moment and place at which communication about the disruption is received (Adele et al., 2019). In response to short-term disruptions, riders are more likely to cancel their trip compared to during long-term disruptions (Nguyen-Phuoc et al., 2018a).

*2.3. Unexplored capacity: On-demand ridesourcing for mobility resilience?*

Ridesourcing offers a potential solution for urban passenger mobility in ordinary travel settings, and a large body of research has investigated its demographic and spatial use. However, little is known about the ability of these on-demand modes like ridesourcing to provide urban mobility resilience during disruptions (Borowski & Stathopoulos, 2020). The following sections summarize what is known about user patterns and social inequities to guide the analysis of the role of ridesourcing platforms in providing mobility redundancy during disruptions.

*2.3.1. Ridesourcing demand user and spatial profiles*

Consistently, studies have shown that ridesourcing users tend to have higher income (Sikder, 2019), be younger, (Clewlow & Mishra, 2017; Rayle et al., 2016; Young & Farber, 2019), highly educated (Alemi et al, 2018; Dias et al., 2017), full-time workers (Shamshiripour et al., 2020), male (Zhang & Zhang, 2018), own fewer vehicles per household, and live closer to transit stations (Deka & Fei, 2019). Looking at spatial use patterns, recent analyses of large-scale ridesourcing trip data reveal that several aggregate city-specific factors correlate with ridesourcing trip counts. Greater ridesourcing usage is shown to be positively correlated with population in Austin (Lavieri et al., 2018), Chicago (Ghaffar et al., 2020), and Los Angeles (Brown, 2019b), employment density in Austin (Lavieri et al., 2018), Chicago (Ghaffar et al., 2020), Los

5Angeles (Brown, 2019b), and New York City (Correa at al., 2017), and land-use diversity in Austin (Yu & Peng, 2019) and Chicago (Ghaffar et al., 2020), as well as lower household income in Los Angeles (Brown, 2019b) and zero-vehicle households and percentage transit commuters in Chicago (Ghaffar et al., 2020).

Relationships of ridesourcing with competing transport options, like public transit, are less clear. Some evidence suggests that public transport demand is reduced by the door-to-door business model of ridesourcing (Clewlow & Mishra, 2017). In Chicago, Marquet (2020) finds that ridesourcing is used to travel between areas that are already highly accessible by transit, suggesting the potential to complement transit due to market density. Soria and Stathopoulos (2020) note that the link between ridesourcing and transit varies across cities and can be either competing or complementary, warranting continued research to pinpoint the evolving and location-specific mode connections. Overall, evidence indicates that ridesourcing is related to rider privilege and that most ridership takes place in the dense and accessible urban core (Lewis & MacKenzie, 2017; Soria & Stathopoulos, 2021).

*2.3.2. Social inequity concerns related to on-demand mobility*

The question of demographic or spatial exclusion related to on-demand mobility platforms has attracted significant attention (Pangbourne, 2020). The ridesourcing business model has been accused of being based on privileged access due to limited accessibility to rural populations, under-banked households, individuals without smartphones, individuals with disabilities, lower income groups, or other historically marginalized communities (Daus Esq., 2016). Additionally, reliance on registration and digitalization may generate accessibility inequalities (Shaheen et al., 2017). Ridesourcing companies continue to grapple with problems of the digital divide, discrimination, data privacy, and workers' rights (Jin et al., 2018).

Prior research has shown evidence that ridesourcing service quality differs according to race, ethnicity, and income, which raises important equity concerns. Ge et al. (2016) points to patterns of discrimination in ridesourcing practices. In Seattle, longer wait times for ridesourcing services at night were observed in areas with higher percentages of racial minorities (Hughes & Mackenzie, 2016). A similar trend of longer ridesourcing wait times has been observed for communities of color in Chicago (CNT, 2019). In New York City, ridesourcing pickup rates were lower in lower income areas (Correa et al., 2017; Jin et al., 2019). By contrast, other recent research instead argues that ridesourcing is closer than traditional taxis to eradicating racial and ethnic inequities in service quality (Brown, 2019a). In Atlanta, estimated wait times of UberX and UberBLACK were not significantly correlated with race or income (Wang & Mu, 2018).

Looking at broad spatial ridership patterns for the case of Chicago, research confirms a greater concentration of rides in the more affluent North and Central districts of the city (Brown, 2019b; Ghaffar et al., 2020; Soria et al., 2020). Fewer ridesourcing trips are generated in areas that are predominantly Hispanic or African American, lower income, or had lower rates of car ownership (Marquet, 2020). Similarly, solo ridesourcing demand is lower in areas where socioeconomic disadvantage is concentrated (Soria & Stathopoulos, 2021). These findings mirror research on another on-demand platform; Divvy bikeshare uptake is lower in the less affluent, majority African American, South district of Chicago (Biehl et al., 2018).

*2.3.3. Linking on-demand mobility to disruptions*

Only recently has research begun to consider the role of ridesourcing in addressing resilience in the context of no-notice transit service disruptions. Borowski and Stathopoulos (2020) found that ridesourcing may be used to connect with transit during no-notice urban evacuations. One study of ridesourcing during unplanned subway disruptions in Toronto suggests less frequent shifts to ridesourcing in disadvantaged neighborhoods and inequitable bus bridging services (Liu et al., 2020). For Chicago, stated preference survey research on ridesourcing use during transit disruptions shows that Millennial status, higher level of education, smartphone access, and prior ridesourcing experience is associated with a shift to ridesourcing (Rahimi et al., 2020). A second stated preference survey shows that the preference for on-demand ridesourcing during no-notice mobility disruptions is shaped by identity factors, such as the intersection of race, gender, and class identities (Borowski & Stathopoulos, 2020).

The present study builds on earlier work investigating the role of bikesharing as a gap-filling mechanism during longer-term transit disruptions wherein a temporary increase in bikeshare demand during transit strikes and maintenance projects was noted, suggesting the ability of on-demand transportation modes to improve mobility resilience (Fuller et al., 2012; Kaviti et al., 2018; Pu et al., 2017; Saberi et al., 2018). In this study, we focus on shifts to ridesourcing during unplanned, short-term transit service disruptions. While similar research has been conducted in Toronto (Liu et al., 2020), our work is among the first to investigate mobility resilience in the context of Chicago with its unique transit and sociodemographic patterns.

### 2.4. *Literature takeaways*

Acute shocks and chronic stressors are likely to continue to worsen in urban areas due to climate change and aging infrastructure. In this context transit agencies face the risk of exacerbated ridership abandonment following unplanned service disruptions. This study examines an untapped potential of emergent, on-demand modes to mitigate the negative impacts of no-notice transportation system shocks, thereby boosting mobility resilience. However, this potential is not without shortcomings. Although interest in equity is growing, its relationship with mobility resilience remains understudied (Mattsson & Jenelius, 2015). Here we carefully consider the equity context of on-demand mobility usage to guide our selection and interpretation of station-, community-, and district-level predictor variables. We use a multilevel mixed (MLM) modeling analysis to examine spatially determined variation in resilience across the city. This research fills current gaps in the literature related to the question of "resilience for whom" while considering multimodal aspects of resilience. In this study, which is among the first to use a natural experiment to examine ridesourcing behavioral responses to no-notice, short-term transit disruptions, we hypothesize that adaptive ridesourcing will be associated with privileged sociodemographics and resource access (Deka & Fei, 2019).

## 3. Data and model

### 3.1. *Case study context*

In this study, we analyze the equity of ridesourcing for mobility resilience in Chicago. Notably, Chicago is home to the second largest transit network in the U.S. with the Chicago Transit Authority (CTA) serving 3.5 million riders (CTA, 2020a) with nearly 16 million rail transit rides each month (CTA, 2020b). While unequal access to essential resources is common in many U.S. cities, Chicago contends with historically rigid, spatially defined, social and economic inequality that is frequently linked to race. For example, the income divide between White households and racial minority households is wider in Chicago than it is across the nation (Asante-Muhammed, 2017). Additionally, urban mobility systems typically contend with multiple layers of inequality of mobility investments and service access that determine service quality for different population segments (Lowe, 2014). Chicago is subject to urban mobility inequities both in terms of service provisions (e.g., poor mobility accessibility and lack of pedestrian-friendly infrastructure) and disproportionate impacts (e,g., biased policing) in low-income communities (Krapp, 2020, Barajas 2021).

**Fig. 1** shows the spatial distributions of community area **(a)** median household income, **(b)** percentage of people of color, **(c)** percentage of transit commuters, **(d)** population density, **(e)** total ridesourcing trips in 2019, and **(f)** percentage of active mode commuters. These heatmaps confirm the following three observations. First, there is a narrowly concentrated demand for ridesourcing and active mobility commuting in the Central district of the city, representing the urban core from which the rail transit network expands radially (**e** and **f**). Second, there is a greater concentration of population and transit commuting in the North district (**d** and **c**). Third, the distribution of income is largely opposite of the majority non-White racial breakdown (**a** and **b**). **Fig. 2** shows the layout of the Chicago Transit Authority rail network for comparison.

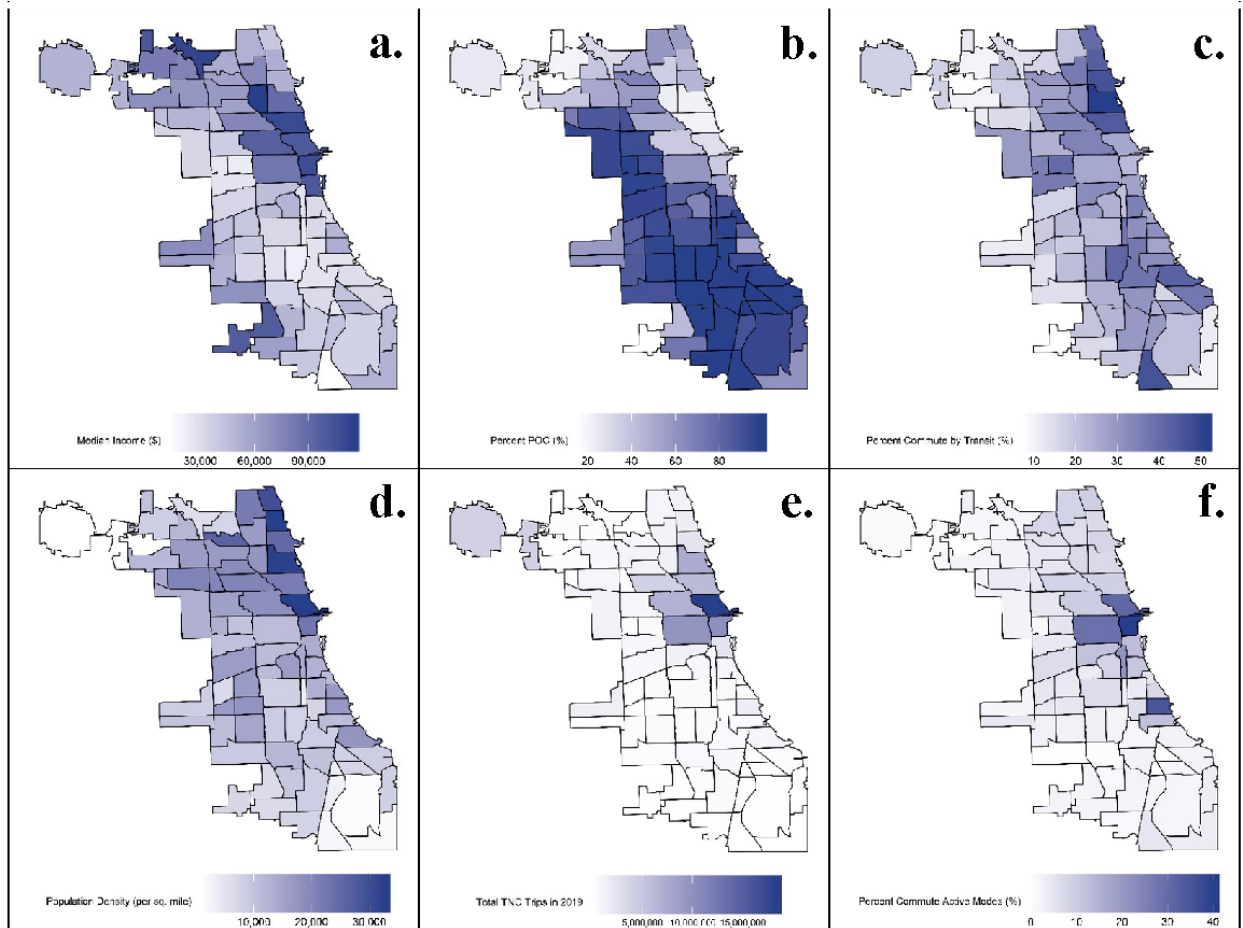

**Fig. 1.** Heatmaps showing the spatial distribution of **(a)** median household income, **(b)** percent population of color, **(c)** percent commute by transit, **(d)** population density, **(e)** total transportation network company ridesourcing trips in 2019, and **(f)** percent commute by active (walk and bike) modes.

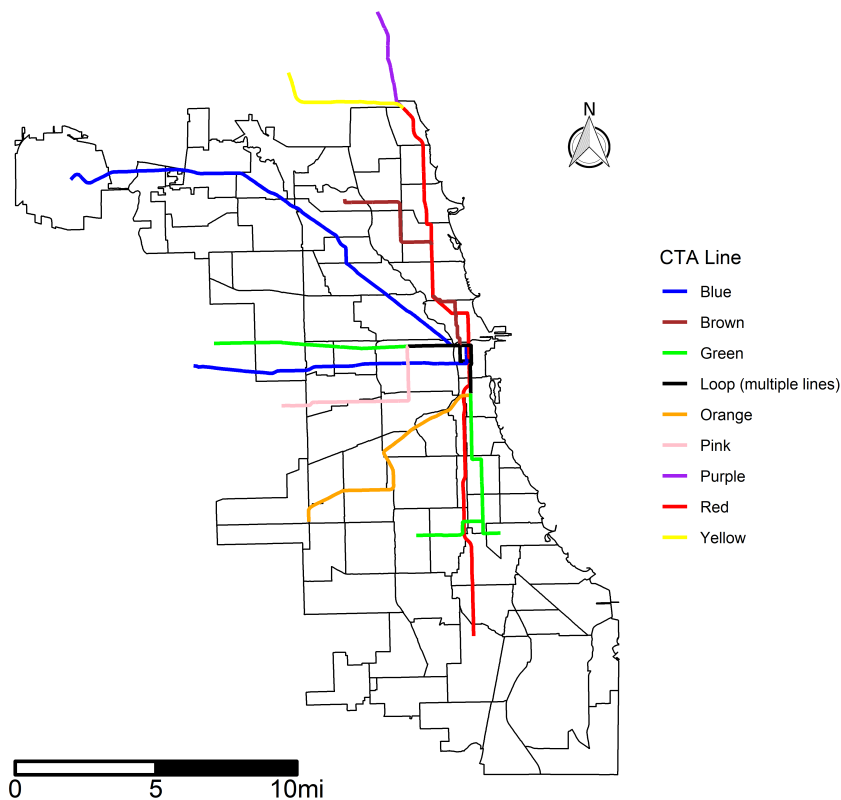

**Fig. 2.** Map of the Chicago Transit Authority rail network.

### *3.2. Data and variable description*

The dataset obtained from the City of Chicago data portal is freely available for download and consists of over 152 million trips by Uber, Lyft, and Via spanning the period of November 2018 through October 2019 (Chicago Data Portal, 2019). **Table 1** lists the variables used in this study to describe ridesourcing trips, disruption events, and community areas.

#### *3.2.1. Community area sociodemographics*

The City of Chicago comprises 77 community areas that can be further aggregated into four districts (i.e., North, Central, South, and West). For reasons of privacy, individual-level data on ridesourcing trip-makers, such as sociodemographics and residential locations of rail and ridesourcing riders, are not publicly available (City of Chicago, 2020). Therefore, we follow the practice of using aggregated measurements of community sociodemographics to represent sociodemographic variables of interest (Liu et al., 2020). While this practice can mask variation in rider characteristics, it is currently the best option available to researchers.

#### *3.2.2. Disruption-based ridesourcing demand*

Twenty-eight CTA rail transit disruptions lasting a minimum of one hour are identified as having occurred during the period of November 2018 through October 2019 using a Google News search for the phrase "CTA disruption". The timespan for our study was truncated to a single year due to the major impacts of the COVID-19 pandemic on public transit and ridesourcing ridership beginning in early 2020. **Fig. 3** highlights the locations of these transit disruption sources at station, community, and district levels, and it can be observed that all CTA lines experienced disruptions during this period. **Table 2** lists the disruption events and identifies their locations, impacted stations, whether the disruption occurred during peak travel hours, and whether a shuttle bus was deployed by CTA to assist riders according to the associated report.

| Variable Name | Minimum | Maximum | Mean | Standard Deviation | Data Source |
|---|---|---|---|---|---|
| Air temperature | 11.00 | 85.45 | 53.83 | 18.10 | (CTA, 2019) |
| Airport indicator | 0.00 | 1.00 | 0.06 | 0.24 | (Google, 2019) |
| Area | 0.71 | 32.47 | 3.83 | 6.31 | (CMAP, 2019) |
| Deployment of shuttle bus | 0.00 | 1.00 | 0.63 | 0.48 | (CTA, 2019) |
| Disruption duration | 1.00 | 7.00 | 2.49 | 1.52 | (CTA, 2019) |
| District (North side) | 0.00 | 1.00 | 0.50 | 0.50 | (CTA, 2019) |
| Holiday indicator | 0.00 | 1.00 | 0.07 | 0.26 | (CTA, 2019) |
| Late night (after 10 PM) indicator | 0.00 | 1.00 | 0.05 | 0.23 | (CTA, 2019) |
| Median household income (in thousands) | 19.80 | 104.35 | 55.44 | 28.40 | (CMAP, 2019) |
| Medical emergency indicator | 0.00 | 1.00 | 0.58 | 0.50 | (CTA, 2019) |
| Number of bus stations | 0.00 | 129.00 | 36.17 | 28.83 | (CTA, 2019) |
| Number of Divvy stations | 0.00 | 30.00 | 4.02 | 5.54 | (Divvy, 2020) |
| Number of ridesourcing trips | 0.00 | 5581.00 | 635.57 | 1001.46 | (Chicago Data Portal, 2019) |
| Number of stations impacted | 1.00 | 15.00 | 8.46 | 4.16 | (CTA, 2019) |
| Peak hour indicator | 0.00 | 1.00 | 0.21 | 0.41 | (CTA, 2019) |
| Percentage of commuters taking transit | 10.80 | 54.20 | 34.26 | 11.78 | (CMAP, 2019) |
| Percentage of residents who self-identify as White non-Hispanic | 0.70 | 81.50 | 38.06 | 30.65 | (CMAP, 2019) |
| Percentage of zero vehicle households | 8.60 | 57.40 | 33.50 | 12.06 | (CMAP, 2019) |
| Population (in thousands) | 2.44 | 100.47 | 40.57 | 25.64 | (CMAP, 2019) |
| Population density (in thousands per square mile) | 0.38 | 32.73 | 16.52 | 9.64 | (CMAP, 2019) |
| Precipitation | 0.00 | 0.04 | 0.00 | 0.01 | (CTA, 2019) |
| Source of disruption | 0.00 | 1.00 | 0.31 | 0.47 | (Google, 2019) |
| Station ridership (in thousands) | 0.65 | 58.05 | 12.26 | 11.39 | (CTA, 2019) |
| Weekday indicator | 0.00 | 1.00 | 0.63 | 0.48 | (CTA, 2019) |

**Table 1.** Variables considered in this study

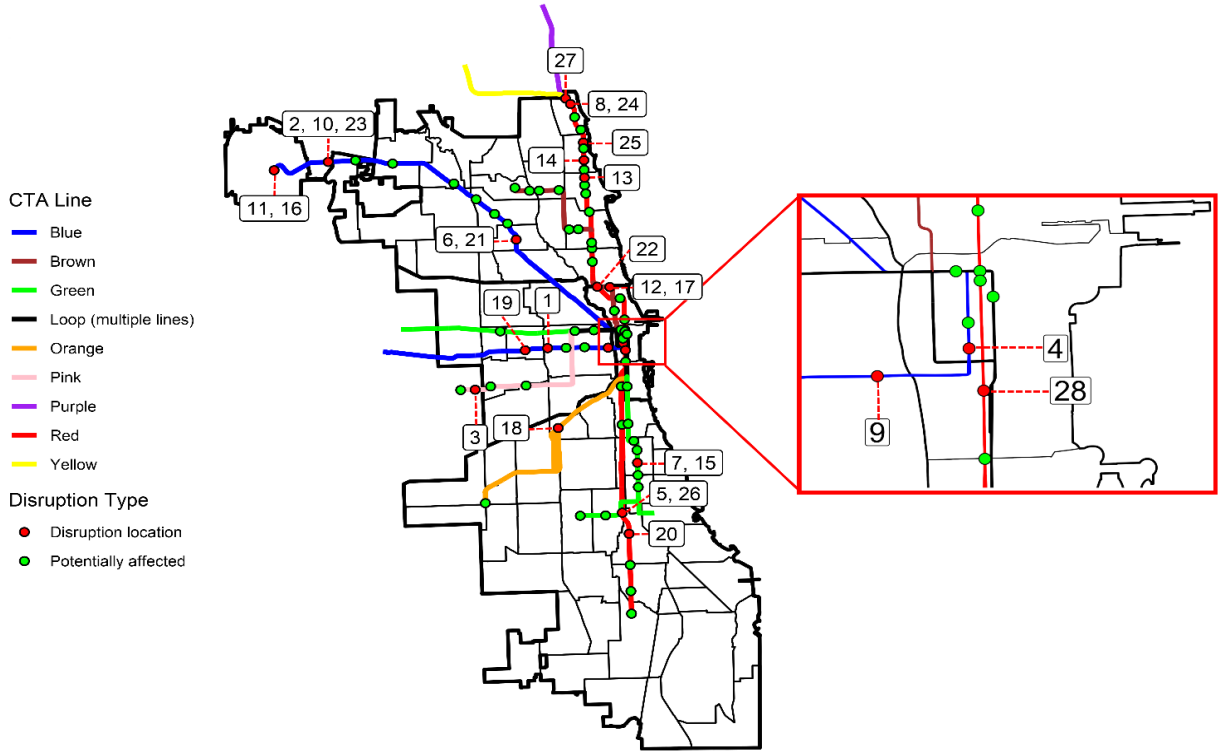

**Fig. 3.** Map of Chicago showing rail transit lines and disruption events within community area and district boundaries. Insert represents the Loop (i.e., the central business district).

*3.2.3. Baseline ridesourcing demand*

To generate a robust four-day ridesourcing demand baseline, trip counts during the disruption period are averaged across the same day of week and time of day (i.e., the specified disruption period) as the disruption for two weeks prior to the event and two weeks following, as in Liu et al. (2020). This was done to account for station accessibility and seasonality. Each ridesourcing trip is included in the analysis if the starting location is within a 0.25-mile radius of a disrupted transit station. This frequently used walking estimate (Younes et al., 2019; Zhao et al., 2003) is applied to account for riders who source rides on their way to or from the impacted transit station, such as to facilitate ridesourcing pick-up by avoiding the potential crowd surrounding the disrupted station.

*3.3. Multilevel mixed model specification*

To address the research question of ridesourcing surges triggered by transit disruptions, we control not only for the immediate station attributes where the disruption occurs, but also for community area and district level factors in a multilevel mixed (MLM) structure as shown in **Fig. 4**. MLM models provide a mechanism for analyzing datasets where events (in this case, station disruptions) are nested within higher-order spatial contexts and correctly account for the hierarchical nesting of data and effects happening at different levels (Goldstein, 2003; Julian, 2001; Wampold & Serlin, 2000). In the past, MLM or hierarchical models have been used to represent the structure of social relations within personal networks (Carrasco & Miller, 2009), temporal changes in bikeshare trips (El-Assi et al., 2017), and transit demand between origin-destination station pairs (Iseki et al., 2018). Here we use the multilevel analysis to identify factors associated with systematic variations in ridesourcing demand during transit disruptions at the station, community area, and city district levels. We can thereby examine explanatory variables at each level of the data hierarchy, and in doing so, control for community area effects on station ridership variation.

| Number | Date | Day | Start Time | End Time | District | Source Station | Impacted Span | Stations Impacted | Peak Hour | Shuttle Bus |
|---|---|---|---|---|---|---|---|---|---|---|
| 1 | 11/06/18 | Tuesday | 5:00 | 6:00 | West | Western | Pulaski to Racine | 5 | | ✓ |
| 2 | 11/12/18 | Monday | 13:30 | 16:30 | North | Rosemont | Harlem to O'Hare | 4 | | ✓ |
| 3 | 11/26/18 | Monday | 9:00 | 12:15 | West | Cicero | 54th/Cermak to Pulaski | 2 | ✓ | |
| 4 | 12/06/18 | Thursday | 17:00 | 18:00 | Central | Jackson | Jackson | 1 | ✓ | |
| 5 | 12/12/18 | Wednesday | 5:00 | 8:30 | South | 63rd | 47th to 95th/Dan Ryan | 7 | | ✓ |
| 6 | 12/17/18 | Monday | 8:00 | 10:00 | North | Belmont | Addison to Fullerton | 5 | ✓ | ✓ |
| 7 | 01/12/19 | Saturday | 12:30 | 14:00 | South | 47th | 63rd to Sox-35th | 4 | | ✓ |
| 8 | 01/20/19 | Sunday | 9:00 | 10:30 | North | Jarvis | Belmont to Howard | 14 | | ✓ |
| 9 | 02/14/19 | Thursday | 13:00 | 16:00 | West | Clinton | Ashland to Washington/Wabash | 6 | | ✓ |
| 10 | 03/12/19 | Thursday | 21:00 | 3:00 | North | Rosemont | Jefferson Park to O'Hare | 5 | | ✓ |
| 11 | 04/10/19 | Wednesday | 19:00 | 2:00 | North | O'Hare | O'Hare to Rosemont | 2 | ✓ | ✓ |
| 12 | 05/01/19 | Wednesday | 7:20 | 8:20 | North | North/Clybourn | Cermak-Chinatown to Fullerton | 5 | ✓ | |
| 13 | 05/06/19 | Monday | 16:00 | 18:00 | North | Argyle | Argyle | 1 | ✓ | |
| 14 | 05/12/19 | Sunday | 14:00 | 16:00 | North | Bryn Mawr | Addison to Howard | 14 | | |
| 15 | 06/06/19 | Thursday | 11:00 | 16:30 | South | 47th | Ashland/63rd to Roosevelt | 10 | | ✓ |
| 16 | 06/10/19 | Monday | 9:00 | 10:00 | North | O'Hare | O'Hare to Rosemont | 2 | ✓ | |
| 17 | 06/12/19 | Wednesday | 19:20 | 20:20 | North | North/Clybourn | Cermak-Chinatown to Fullerton | 11 | ✓ | |
| 18 | 06/20/19 | Thursday | 10:15 | 13:30 | South | 35th/Archer | Halsted to Midway | 7 | | ✓ |
| 19 | 06/25/19 | Tuesday | 7:30 | 8:45 | West | Kedzie-Homan | Kedzie-Homan | 1 | ✓ | |
| 20 | 06/27/19 | Thursday | 12:30 | 15:00 | South | 69th | 63rd to 95th/Dan Ryan | 5 | | ✓ |
| 21 | 09/07/19 | Saturday | 14:00 | 15:15 | North | Belmont | Fullerton to Kimball | 15 | | |
| 22 | 09/24/19 | Tuesday | 9:00 | 10:30 | North | Sedgwick | Sedgwick | 1 | ✓ | |
| 23 | 09/26/19 | Thursday | 1:00 | 4:00 | North | Rosemont | Harlem to O'Hare | 4 | | |
| 24 | 09/26/19 | Thursday | 17:45 | 22:00 | North | Jarvis | Belmont to Howard | 14 | ✓ | |
| 25 | 10/05/19 | Saturday | 22:45 | 2:15 | North | Granville | Belmont to Howard | 14 | | |
| 26 | 10/08/19 | Tuesday | 15:15 | 16:15 | South | 63rd | Roosevelt to 95th/Dan Ryan | 10 | | ✓ |
| 27 | 10/30/19 | Wednesday | 15:00 | 16:00 | North | Howard | Belmont to Howard | 14 | | |
| 28 | 10/31/19 | Thursday | 16:15 | 18:30 | Central | Harrison | Cermak-Chinatown to Fullerton | 11 | | ✓ |

**Table 2.** Twenty-eight unplanned rail transit disruptions in Chicago (Nov. 2018 - Oct. 2019)

The advantage of using the multilevel structure is the ability to estimate the variability in results that can be attributed to neighborhood (e.g., community area) effects rather than only to individual station effects. By carefully controlling variable-inclusion at the appropriate level, the model considers correlations between observations within the same group (i.e., a given community area) as distinct from correlations between groups (Jones & Duncan, 1996). In contrast, a standard one-level regression model would ignore group-level distinctions (e.g., different commuting patterns in different communities) and group-level correlations (e.g., similar patterns of use among stations in the same community related to the income-level of riders). A useful way to think of MLM models is as a structure positioned between two modeling extremes when groupings are known: fully pooled and fully unpooled specification (Gelman & Hill, 2007). A fully pooled model treats group-level variables as individual variables, thereby ignoring group-level distinctions. The opposite extreme, a fully unpooled model, asserts that the groups are so completely different that they cannot be associated in the same model. The MLM model offers a compromise between these perspectives by modeling individual-level fixed effects as well as distributional assumptions on the random effects.

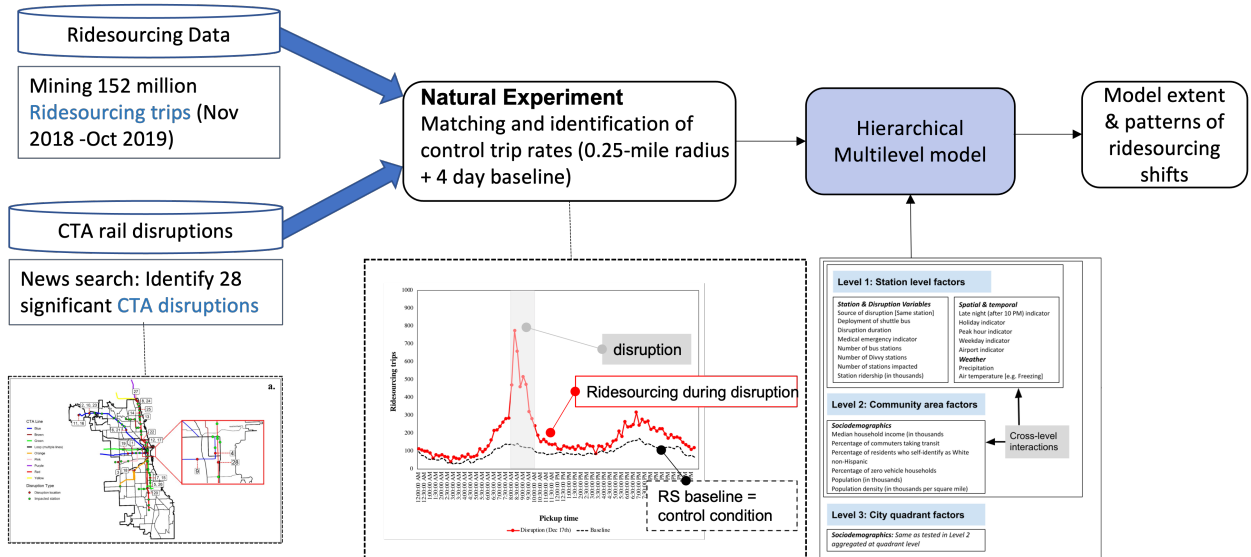

**Fig. 4.** Framework for the hierarchical multilevel mixed model of ridesourcing shifts during transit disruptions. Station-level disruptions are nested within community areas, which in turn are nested within districts.

**Fig. 5** shows each variable tested in the modeling along with the inclusion strategy for each level of analysis. The dependent variable is the number of ridesourcing trips compared to the baseline demand two weeks prior and two weeks following the disruption (i.e., individual station observations). Covariates related to the disruption cause, context, and timing are included as explanatory variables at this level, in line with Mojica (2008) and Pu et al. (2017). We further investigate whether the fact that stations are nested within community areas and major districts plays a role in ridesourcing demand shifts. It is likely that a comparable disruption can generate different mode-shifting effects depending on where it is located, owing to the different composition of travelers and availability of alternative modes. Specifically, the broader context is controlled for by including sociodemographic and mobility factors measured at the community level, which are in turn aggregated to the district level of analysis. We apply group mean centering for community area variables (Enders & Tofighi, 2007) to facilitate the interpretation of the cross-level interactions. It is worth noting that since the disruptions we measure result from a natural experiment, we are unable to control exhaustively for all combinations of factors that are at play within and between community areas. Therefore, we include random intercept effects at each of the lower-nested group levels to partition the unexplained variability effects on the dependent variable.

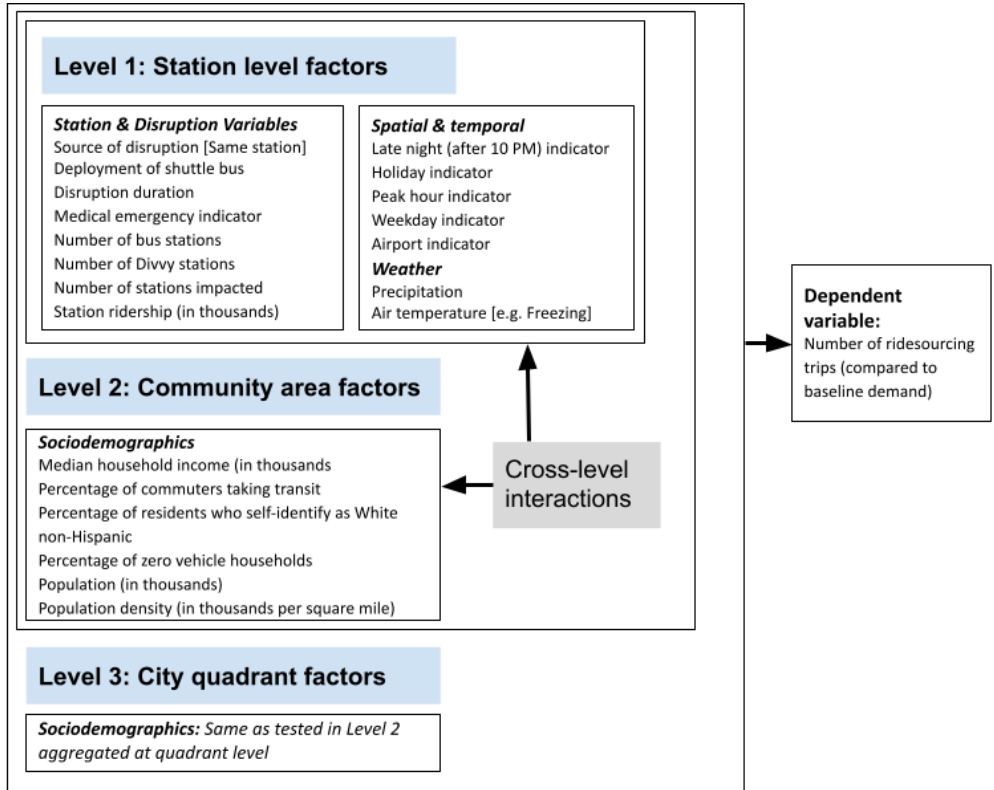

**Fig. 5.** Multilevel model variables. *Considered variables are listed for each level.*

Conceptually, the model can be articulated as regression equations occurring at different levels where each group-level coefficient has its own regression equation. Following Gill and Womack (2013), the general three-level structure is defined in **Eq. 1** as:

$$y_{ijk} = \beta_{0jk} + \beta_{1jk} x_{1ijk} + \varepsilon_{ijk} \tag{1}$$

where *i* represents the station, *j* represents the community area, and *k* represents the district. $\beta_{0jk}$ is the (random) intercept measuring average ridesourcing use (defined in **Eq. 2**), and $x_{1ijk}$ is a predictor, such as the average daily transit use measured at the station level, while $\beta_{1jk}$ is the (random) slope depicting the relationship between the station-level variables and the change in ridesourcing demand (as defined in **Eq. 3**). The error term $\varepsilon_{ijk}$ relates to station-level effects.

By including Level 2 and 3 explanatory variables in the model, we uncover broader area effects. The Level 2 formulation includes variables aggregated to the community area level. This can be thought of as being equivalent to the way in which student educational performance is affected by their classroom teacher in a way that is distinct from the effects of their individual factors or from more aggregate school-level effects. At Level 2, the general regression equations are defined as:

$$\beta_{0jk} = \gamma_{00k} + \gamma_{01k} x_{2jk} + u_{0jk} \tag{2}$$
$$\beta_{1jk} = \gamma_{10k} + \gamma_{11k} x_{2jk} + u_{1jk} \tag{3}$$

where the random intercept $\beta_{0jk}$ is a function of $\gamma_{00k}$, which is the grand mean of ridesourcing demand surges across stations in the community (defined below in **Eq. 4**). The subscript *jk* denotes the distinct community area impacts. The ɣ random effects coefficient has numbered subscripts, the first denotes the intercept (0) or slope (1), while the second subscript denotes the independent variable. Departures from this average intercept represented by $x_{2jk}$ are community-level predictors with $\gamma_{01k}$ denoting the random slope for community-level predictors (**Eq. 5**), and $u_{0jk}$ is the unique effect associated with communities assumed to have a multivariate normal distribution. The random slope $\beta_{1jk}$ is a function of $\gamma_{10k}$ representing the average effect of the station-level predictors (i.e., the slope over all stations shown in (**Eq. 6**)). Departures from the slope (i.e., random effects) over station predictors are represented by the $\gamma_{11k}$ coefficient (**Eq. 7**) that would be removed in the case of a random intercept-only model (as in the current analysis).

At Level 3, variables vary by district and apply to all individual cases and community areas assigned to this group. Therefore, they contain the subscript *k* as opposed to *ijk* or *jk*. At Level 3, the separate regression equations for the intercepts and slopes are defined as:

$$\gamma_{00k} = \delta_0 + \delta_4 x_{3k} + u_{00k} \tag{4}$$
$$\gamma_{01k} = \delta_2 + \delta_5 x_{3k} + u_{01k} \tag{5}$$

$$\gamma_{10k} = \delta_1 + \delta_6 x_{3k} + u_{10k} \tag{6}$$
$$\gamma_{11k} = \delta_3 + \delta_7 x_{3k} + u_{11k} \tag{7}$$

where $\delta_0$ is the intercept shared by all individual cases, $\delta_1, \delta_2,$ and $\delta_3$ are the main effects, $\delta_4, \delta_5,$ and $\delta_6$ are two-way interactions, and $\delta_7$ is a three-way interaction.

In our specific modeling, the outcome variable of the three-level hierarchy $y_{ijk}$ is defined as the change in ridership over baseline. After specification testing, the final model takes the specific form as shown in **Eq. 8-16**. The model includes a random intercept $\beta_{0jk}$ and two main effects ($non\_holiday_{ijk}$ and $peak\_hour_{ijk}$) at Level 1, shown in **Eq. 8**. Level 2 brings in contextual variables used to explain variability in ridesourcing demand via cross-level interactions. That is, we model the intercept and slopes explicitly, and include Level 1 and Level 2 independent variables interacted to describe variation in the intercept. **Eq. 9-11** show the random intercept $\gamma_{00k}$ and the cross-level interaction terms ($percent\_white_{jk} \times peak\_hour_{ijk}$ and $percent\_transit_{jk} \times disruption\_source_{ijk}$). Level 2 also specifies $\beta_{1jk}$ and $\beta_{2jk}$ which represent the parameter slopes with $\gamma_{10k}$ and $\gamma_{20k}$. Level 3 includes the random intercept $\delta_0$ and one district-level interaction ($north\_quad_k \times shuttle_{ijk}$) that is found to generate variability in ridesourcing (**Eq. 12**), with remaining parameters $\delta_1$ and $\delta_2$ denoting the fixed slope coefficients. The disturbance parameters are included at the community $u_{0jk}$ and district levels $u_{00k}$ (**Eq. 15-16**).

<div align="center">Level 1 Model: Station Effects</div>

$$y_{ijk} = \beta_{0jk} + \beta_{01k,non\_holiday}\, non\_holiday_{ijk} + \beta_{peak\_hour} peak\_hour_{ijk}\, peak\_hour_{ijk} + \varepsilon_{ijk} \tag{8}$$

<div align="center">Level 2 Model: Random Intercept & Cross-Level Interactions at Community Level</div>

$$\beta_{0jk} = \gamma_{00k} + \gamma_{01k}\, x_{percent\ white,jk} \times x_{peak-hour,ijk}$$
$$+ \gamma_{02},\, x_{percent_{transit},jk} \times x_{disruption\_source_{ijk}} + u_{0jk} \tag{9}$$
$$\beta_{1jk} = \gamma_{10k} \tag{10}$$
$$\beta_{2jk} = \gamma_{20k} \tag{11}$$

<div align="center">Level 3 Model: Random Intercept & Cross-Level Interaction at District Level</div>

$$\gamma_{00k} = \delta_0 + \delta_{NorthShuttle}\, north\_quad_k \times shuttle_{ijk} + u_{00k} \tag{12}$$
$$\gamma_{10k} = \delta_1 \tag{13}$$
$$\gamma_{20k} = \delta_2 \tag{14}$$
$$u_{0jk} \sim \mathcal{N}(0, \sigma_d^2) \tag{15}$$
$$u_{00k} \sim \mathcal{N}(0, \sigma_e^2) \tag{16}$$

It is important to note that the cross-level interactions explain a significant amount of variance of ridesourcing demand changes in addition to that already explained by the station-level equations.

## 4. Results and discussion

### *4.1. Neighborhood differences in adaptive ridesourcing during disruptions: Descriptive analysis*

Exploratory analysis shows that adaptive ridesourcing response is not uniform across the city. Two different poles are exemplified in Fig. 6. **Fig. 6(a)** shows a significant surge in the use of ridesourcing following a no-notice rail transit disruption. This high-impact North district case in Lake View at the Belmont station (the source of the disruption) occurred on a Monday in December during morning peak hours and was caused by a train striking a person. The baseline ridesourcing demand for this time and location is 807 rides, meaning the disruption is associated with a statistically significant surge in ridesourcing trips totaling 2,883, which corresponds to a 257% increase.

**Fig. 6(b)** shows a similar disruption event occurring in an under-resourced West district neighborhood with limited shifting to on-demand services. This low-impact disruption in East Garfield Park at the Kedzie station (the source of the disruption) resembles the Belmont disruption in that it occurred during weekday morning peak hours and was caused by a person on the tracks. However, the baseline ridesourcing demand for this time and location is a fraction of that at Belmont: only 89 rides. The number of ridesourcing rides during the disruption event is lower than the baseline of 76 (an insignificant decrease of 15%). This observed difference in disruption response is likely related to more pervasive racial and economic inequities, along with differences in transit accessibility. Specifically, Lake View has a median household income of $86,119 and 79% of its residents are White, while East Garfield Park has a median household income of $23,116 and 5.6% of its residents are White. To systematically examine different patterns of ridesourcing demand shifts prompted by transit disruptions across Chicago, we turn to our MLM model results.

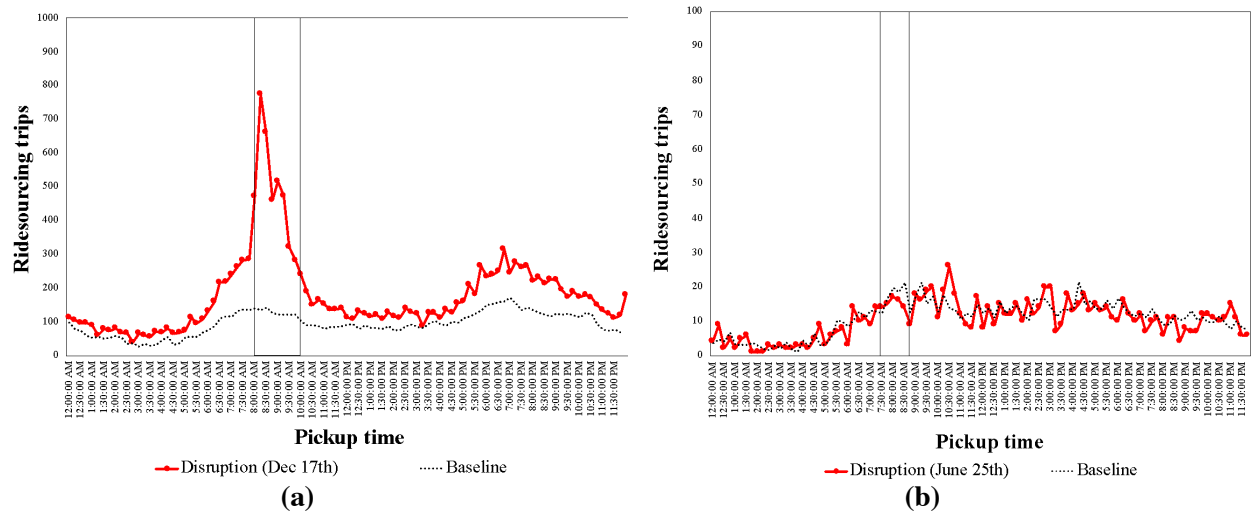

**Fig. 6** Ridesourcing trips during disruptions (solid line) compared to baseline (dotted line) at **(a)** the Belmont station in the North district and **(b)** the Kedzie station in the West district. Disruption duration is indicated by a border. The y-axes are scaled according to baseline ridesourcing trips (10:1).

### *4.2. Empirical model specification*

Three models are estimated: (1) a basic intercept model, (2) a station-level analysis, and (3) a model accounting for cross-level effects. For model building, each of the hypothesized predictors measured at the station level is tested first independently and then jointly. To model explanatory variables, we follow the block entry approach consisting of the gradual addition of covariates level by level (Cohen et al., 1983), following the plan outlined in **Fig. 5**. Data preparation and merging was done using R 3.5.0 (R Core Team, 2020), with modelling done in Stata using the *mixed* function for multilevel mixed-effects linear regression (StataCorp, 2019).

**(1) Basic model specification**

The results of three MLM models are shown in **Table 3**. These models include significant fixed effects and explanatory features, and all parameters are significant to a 98.9% level of confidence or greater except for the model constants. Station-level (Level 1) fixed effects (i.e., *non-holiday disruption* and *peak hours disruption*) resemble standard regression parameters. Three explanatory features reflect the context surrounding the station, namely two cross-level (station- and community-level) random effects (i.e., *percent White during peak hours* and *percent transit commuters at the source of disruption*) and one district-level effect (i.e., *shuttle deployment in the North district*).

The empty MLM model (Model 1) partitions the variance at each aggregation level without including any explanatory variables. This null model is used to calculate the intraclass correlation (ICC), also known as the variance partition coefficient, for the three levels of analysis (Snijders & Boskers, 1999). Thereby, Model 1 provides an estimate of a baseline variance of ridesourcing demand shifts attributed to factors beyond the immediate station (i.e., community- and district-level factors). The intra-community correlation reveals that the largest share of variation in ridesourcing demand (43%) is related to community-level factors while the station level explains 37% of the variance.

|  | Model 1 | | | Model 2 | | | Model 3 | | |
|---|---|---|---|---|---|---|---|---|---|
| **Fixed part** | Coef. | z value | P > \|z\| | Coef. | z value | P > \|z\| | Coef. | z value | P > \|z\| |
| Non-holiday |  |  |  | 541 | 2.97 | 0.003 | 599 | 3.58 | 0.000 |
| Peak hour |  |  |  | 408 | 4.11 | 0.000 | 353 | 4.22 | 0.000 |
| Constant | -53.2 | -0.39 | 0.700 | 54.1 | 0.97 | 0.332 | 91.8 | 1.53 | 0.125 |
| *Cross-level interactions* |  |  |  |  |  |  |  |  |  |
| Peak hour * Percent White |  |  |  |  |  |  | 12.3 | 2.69 | 0.007 |
| Disruption source * Percent transit commuters |  |  |  |  |  |  | 18.4 | 2.54 | 0.011 |
| *District-level interactions* |  |  |  |  |  |  |  |  |  |
| North district * Shuttle |  |  |  |  |  |  | 321 | 3.66 | 0.000 |
| **Fit statistics** |  |  |  |  |  |  |  |  |  |
| Log likelihood | -836.99148 |  |  | -822.09863 |  |  | -809.57185 |  |  |
| LR test | 12.17 |  |  | 7.15 |  |  | 17.1 |  |  |
| Prob > χ2 | 0.0023 |  |  | 0.028 |  |  | 0.0002 |  |  |
| AIC | 1681.983 |  |  | 1672.508 |  |  | 1637.144 |  |  |
| R2 |  |  |  | 0.34 |  |  | 0.33 |  |  |
| *Intraclass correlation* |  |  |  |  |  |  |  |  |  |
| Level 3: District | 20% |  |  | 0% |  |  | 0% |  |  |
| Level 2: Community area | 43% |  |  | 56% |  |  | 73% |  |  |
| Level 1: Station | 37% |  |  | 44% |  |  | 27% |  |  |

**Table 3.** Multilevel mixed model results

1717**(2) Model with station effects**

In Model 2, with station-level variables, the *district* random intercept (and thereby ICC) is shown to be insignificant, while the variance is partitioned between the station (44%) and community area levels (56%). Owing to high variable collinearity, only two fixed-effect explanatory variables related to the timing of the disruption and a constant are included in Model 2. These statistically significant effects result in a significant improvement in model fit as measured by the deviance difference (836.99 - 822.10 = 14.89, exceeding the critical $\chi^2$ of 5.99 with alpha set at .05) and AIC reduction.

**(3) Model with community effects**

Along the same lines, in Model 3, when adding cross-level effects by including variables measured at the community area level, the variance explained clearly shifts toward the community area variables. Despite the Level 3 district random intercept collapsing to zero, removing this variance component from the analysis causes a significant reduction in overall model fit.

The main takeaway from the variance controls is that factors occurring across different *community areas* are the most decisive in shaping ridesourcing demand shifts during rail transit disruptions. In other words, the ICC calculation shows the community area context is the main source of differences in ridership shifting strategies. We interpret this to mean that there are significant latent neighborhood effects influencing adaptive ridesourcing behavior. These neighborhood effects likely vary as a function of community culture related to car, transit, and ridesourcing ridership, socioeconomic and political factors, and transportation agency strategies.

*4.3. Model results*

**(1) Station level analysis: Local effects of disruptions**

The model constant suggests a moderate average increase of 54 ridesourcing trips (or 15.6%) during a transit disruption, compared to baseline. To contextualize this finding, the average baseline ridesourcing ridership is 347 trips across the Chicago community areas covered in the disruption analysis. This value represents the ridesourcing demand that would occur for the same station and timespan without the disruption. With this baseline in mind, the timing of the disruption is revealed to be highly impactful. On average, when a disruption occurs on a *weekday* (excluding holidays), ridesourcing trips increase by 541 from baseline (a 156% increase). When a transit disruption occurs during *peak hours*, ridesourcing demand increases by 408 rides from baseline (a 118% surge).

These observed citywide trends are likely related to the less flexible trips that occur during peak hours and weekdays. This is not surprising considering that business and commuting trips are more likely to be shifted to another mode than canceled, as shown for planned disruptions (Van Exel & Rietveld, 2009) and for unreliable metro services (Pnevmatikou & Karlaftis, 2011). Our findings for on-demand ridesourcing shifts are novel given that previous research in this area has been dominated by car substitution and bus replacement.

**(2) Community context effects**

In Model 3, we examine the role that the disrupted station's surrounding context plays in determining the transfer of ridership from transit to ridesourcing during no-notice disruptions. Model 3 reveals a significant impact of two community area level factors: *racial composition* and *percent transit commuters*. The addition of these cross-level factors leads to significant improvements in goodness-of-fit measured by the deviance difference and AIC. The positive effect on the interaction term for *percentage of White residents* in the community area with a dummy variable for *peak-hour travel* (a coefficient of 12.3 additional trips) suggests an added effect of racial composition in the local area on the previous peak-hour effect findings. Namely, the peak-hour impact (353 added trips) is amplified when disruptions occur in communities with higher shares of White residents. The implied effect is that a disruption occurring in a community area with a 1% higher share of White residents would result in a boost of 120 (or 3.46%) ridesourcing trips compared to the average peak-hour baseline. This finding adds to existing evidence that ridesourcing provides greater benefits to privileged user groups (Zhang & Zhang, 2018). Given that communities of color in Chicago are more likely to be underserved in ~~relation to~~ job accessibility, transit supply, and on-demand mobility access, we believe this finding is likely a reflection of gaps in access to resources in areas with lower shares of White residents rather than of a lower willingness to use ridesourcing during disruptions, but further research is warranted to gain deeper understanding.

Additionally, a novel effect is found related to the *proportion of transit commuters* in the community area and the *disruption source*. Overall, every percentage unit increase in transit commuting in the community area results in 18 additional ridesourcing trips (or a 5.2% increase). However, this effect is only observed at the station where the incident causing the disruption occurred. We speculate that transit commuters more readily shift to ridesourcing services when they experience and receive information about the disruption *directly*. In other words, riders at the source of the disruption are likely to have more information regarding the nature of the disruptive event (e.g., from official sources and other riders), which will likely factor into their travel adaptation strategy. In areas with less transit commuting, we speculate that there is presumably less *collective* experience with transit disruptions and therefore a higher likelihood of shifting to private modes due to limited opportunities for word-of-mouth information sharing and social influence and thereby greater individualization of adaptive mobility strategies.

Despite a lack of *unexplained* systematic differences related to the district level beyond Model 1, a model search was conducted to explore additional impactful cross-level interactions that incorporate indicator variables for the four districts of the city. The resultant model suggests a unexpected finding. In the *North* district, when a *shuttle bus* is deployed, ridesourcing trips increase by 321 instances (or 92.5%) from baseline. The deployment of replacement bus services for added transit capacity to assist riders during rail disruptions is a common agency response (Pender et al., 2013), but there appears to be an unanticipated (although not surprising) effect of this strategy: a *boost* in ridesourcing requests. We interpret this unexpected increase in ridesourcing to be related to the signaling effect of this action, namely riders could perceive bus deployment as a strong cue for the severity of the disruption and thus its expected duration. For context, the North district is home to the largest share of disruptions in our dataset (16 of 28 or 57%). The North district maintains heavy transit demand by commuters (shown previously in **Fig 1.d**), as well as higher income levels among these commuters (**Fig 1.b**), factors which likely contribute to a greater shift toward ridesourcing in this area.

## 5. Discussion and implications: Collaborations for resilient mobility

The findings in this analysis show that sudden disruptions in urban rail transit are often accompanied by a surge in demand for ridesourcing. This spontaneous mode-shifting behavior invites more work to consider mobility resilience collaborations. If transit operators like the CTA were to adopt a mandate to deliver stop-to-stop service – also known as *mobility as a service* – in the face of unplanned disruptions, they would internalize the responsibility of providing transportation alternatives when a disruption occurs. By communicating the nature of the disruption and anticipated needs, transit agencies could engage – and perhaps support – ridesourcing companies in providing adaptive, gap-filling services to address no-notice disruptions and thereby reduce the adverse impacts experienced by transit riders. The flexibility of ridesourcing services offers on-call availability to provide extra capacity, while shuttle buses can provide services on fixed routes for long-duration disruptions. Ridesourcing companies and drivers could work in tandem with route-around bus services to meet spikes in demand and avoid exploiting the situation through surge pricing. In some cases, it might be more cost efficient for transit agencies to subsidize a shared or pooled ridesourcing service rather than providing shuttle buses, which require equipment maintenance by the CTA and the staffing of short-notice drivers (Shared-Use Mobility Center, 2020).

Such collaborations would need to be in place ahead of no-notice disruptions. Some relevant insights can be gained from looking at the long term market disruptions observed during the COVID-19 pandemic. As an example, the LA Metro in the Los Angeles region was able to leverage a preexisting collaboration with the ridesourcing company Via by expanding their role from providing first- and last-mile services to private, point-to-point trips to accommodate essential travel during the pandemic (Grossman, 2020). This exemplifies the ability of public-private partnerships to increase mobility resilience to unplanned disruptions.

While collaborations between transit providers and ridesourcing companies may provide a way to decrease disruption response time and assist a greater number of affected travelers, these collaborations are not without challenges. The integration of multiple modes for seamless adaptation is difficult due to: (1) contrasting expectations and standards of operators, (2) conflicting operator goals, (3) differences in regulation and standards for operators, and (4) user barriers, including difficulty for some demographics to navigate integrated mobility systems (Butler et al., 2021). Notably, transit providers are required to ensure fair service to all individuals in accordance with Title VI and the Americans with Disabilities Act, while ridesourcing services are not currently held to the same standards.

## 6. Conclusions

### 6.1. Summary of findings

Given the current climate crisis and urbanization, both acute shocks and chronic stressors of all kinds are multiplying in cities, and disruptions are occurring with increased frequency and severity. This study examines the effects of no-notice rail transit disruptions on mode-shifting strategies. Specifically, we examine the role of ridesourcing as an adaptive substitution strategy to fill gaps created by rail disruptions. This study uses a natural experiment to systematically identify and then temporally and spatially match major transit disruptions with ridesourcing trip data for the City of Chicago. A MLM model is used, where the multilevel structure is designed to account for variation in rail-to-transit shifts and to identify whether determinants are local or occurring due to neighborhood differences.

The analysis yields the following main findings and implications:

(1) Following a sudden rail transit disruption, there is evidence of significant localized surges in ridesourcing demand, highlighting that there is spontaneous mobility resilience in the system. The observed demand substitution is strongest during peak-hour and weekday travel, suggesting that ridesourcing provides selective mobility redundancy in relation to mandatory and time-sensitive travel.

(2) Characteristics of the community area where the transit disruption is located are responsible for most of the variation in observed ridesourcing substitution. Greater shifts to ridesourcing occur in community areas that have higher percentages of White residents, especially during peak-hour disruptions, suggesting potential spatial inequities in the capacity for mobility adaptiveness and thus community resilience.

(3) To address negative impacts of transit disruptions on ridership, transit agencies may consider investing in models of mobility as a service. If transit operators like the CTA were to adopt a policy of delivering end-to-end service despite unplanned disruptions, they would maintain responsibility for providing transportation alternatives when service disruptions occur. Our research suggests a potential role for on-demand ridesourcing to address transit service disruptions.

### 6.2. Limitations

Some caveats warrant discussion. First, modeling was based on identification of transit disruptions and shuttle bus deployment gathered from a systematic search of local news sources. As such, these data were aggregated at the station level, and it was assumed that a given disruption lasted the same duration at every impacted station. Second, our approach to analyzing mode-switching behavior was based on a spatial delimitation of a 0.25-mile radius around each impacted station, but mode-shifting behaviors may have occurred across a broader time-space domain, including travelers who learned of the disruption prior to departure. Third, sociodemographics were spatially aggregated, which may mask individual-level rider characteristics. Since individual ridership data were not available, we could not analyze multimodal adaptive strategies for individual travelers.

Despite these limitations, based on conversations with a CTA rail transit agency professional and considering challenges of data availability and accessibility, our method of data aggregation was the best option available to us. Acknowledging the limitations associated with the use of a natural experiment, our research contributes new insights that would be difficult to gage using small-scale stated response data. Specifically, we capture the circumstances of disruptions that lead to real-world shifts to ridesourcing. Thereby the findings of this study shed light on which communities effectively shift to adaptive on-demand mobility during a disruption and which communities must rely on other alternatives.

### 6.3. Future work

Based on our findings, we suggest two main avenues for future research. First, to address the outlined limitations of spatiotemporal data, further collaborative research should aim for a more nuanced analysis of transit riders' behavioral adaptations to better understand socioeconomic determinants of mobility resilience. For example, matching individual-level ridership data by ridesourcing pickup locations with spatiotemporal bus and rail ridership data would reveal more detailed insights into individual user multimodal adaptive mobility strategies.

Second, we encourage the expansion of this investigation using more qualitative analysis. There is a need for further understanding of the adaptative decision-making process that riders use to cope with unplanned travel disruptions. This includes better pinpointing of risk perceptions, communication about disruptions, circumstances of travel, and attitudes related to emerging ridesourcing options. Specifically, by using latent variable modeling we can better capture rider perceptions surrounding the use of ridesourcing as an adaptive mobility strategy, and enable more tailored transportation policies to foster equitable disruption recovery.

.
**Funding acknowledgement**

Removed for double blind review

**References**

Adams, T. M., Bekkem, K. R., & Toledo-Durán, E. J. (2012). Freight resilience measures. *Journal of Transportation Engineering*, 138(11), 1403–1409. doi: 10.1061/(ASCE)TE.1943-5436.0000415

Adelé, S., Tréfond-Alexandre, S., Dionisio, C., & Hoyau, P. A. (2019). Exploring the behavior of suburban train users in the event of disruptions. *Transportation Research Part F: Traffic Psychology and Behaviour*, 65, 344-362.

Alemi, F., Circella, G., Handy, S., & Mokhtarian, P. (2018). What influences travelers to use Uber? Exploring the factors affecting the adoption of on-demand ride services in California. *Travel Behaviour and Society*, 13, 88-104.

Arslan Asim, M., Weiss, A., Kattan, L., & Wirasinghe, S. C. (2021). Transit Users' Mode Choice Behavior During Light Rail Transit Short-Term Planned Service Disruption. *Transportation Research Record*, 03611981211012421.

Asante-Muhammed, D. (2017). The Racial Wealth Divide in Chicago. *Prosperity Now*. Available at: https://prosperitynow.org/files/resources/Racial_Wealth_Divide_in_Chicago_OptimizedforScreenReaders.pdf

Aydin, N.Y., Duzgun, S., & Heinimann, H. (2017). Resilience evaluation for transportation networks accessibility under seismic risk. In: *CUPUM Conference on Computers in Urban Planning and Urban Management*

Baker, D. M., Lopez, E., & Greenlee, A. J. (2021). Transit development and housing displacement: The case of the Chicago Red Line Extension. *Cities*, 115, 103212.

Barajas, J.M., (2021). Biking where Black: Connecting transportation planning and infrastructure to disproportionate policing. Transportation research part D: transport and environment, 99, p.103027.

Ben-Elia, E., Di Pace, R., Bifulco, G.N., & Shiftan, Y. (2013). The impact of travel information's accuracy on route-choice. *Transportation Research Part C: Emerging Technologies*, 26, pp.146-159.

Biehl, A., Ermagun, A., & Stathopoulos, A. (2018). Community mobility MAUP-ing: A socio-spatial investigation of bikeshare demand in Chicago. *Journal of Transport Geography*, 66, 80-90.

Borowski, E., & Stathopoulos, A. (2020). On-demand ridesourcing for urban emergency evacuation events: An exploration of message content, emotionality, and intersectionality. *International Journal of Disaster Risk Reduction*.

Brown, A. E. (2019a). Prevalence and Mechanisms of Discrimination: Evidence from the Ride-Hail and Taxi Industries. *Journal of Planning Education and Research*, 0739456X19871687.

Brown, A. (2019b). Redefining car access: Ride-hail travel and use in Los Angeles. *Journal of the American Planning Association*, *85*(2), 83-95.

Butler, L., Yigitcanlar, T., & Paz, A. (2020). Barriers and risks of Mobility-as-a-Service (MaaS) adoption in cities: A systematic review of the literature. *Cities*, 103036.

Caprì, S., Ignaccolo, M., Inturri, G., & Le Pira, M. (2016). Green walking networks for climate change adaptation. *Transp. Res. Part D Transp. Environ*. 45, 84–95. https://doi.org/10.1016/j.trd.2015.08.005.

Carrasco, J. A., & Miller, E. J. (2009). The social dimension in action: A multilevel, personal networks model of social activity frequency between individuals. *Transportation Research Part A: Policy and Practice*, 43(1), 90-104.

Chan, R., Schofer, J. L. (2016). Measuring transportation system resilience: response of rail transit to weather disruptions. *Nat. Hazards Rev*. 17, 05015004. https://doi.org/10.1061/(ASCE)NH.1527-6996.0000200.

Chicago Data Portal (2019). Transportation Network Provider – Trips. Available at: https://data.cityofchicago.org/Transportation/Transportation-Network-Providers-Trips/m6dm-c72p

City of Chicago (2020). *TNP Reporting Manual*. Available at: https://chicago.github.io/tnp-reporting-manual/

Clewlow, R. R., & Mishra, G. S. (2017). Disruptive transportation: The adoption, utilization, and impacts of ride-hailing in the United States.

CMAP (2018). On To 2050: Comprehensive Regional Plan. Available at: https://www.cmap.illinois.gov/documents/10180/901373/Revised+ON+TO+2050+Draft+Comprehensive+Plan+9-15-18.pdf/dc978895-1a51-7140-48ca-a19b4847971c

CMAP (2019). Community Data Snapshot. Available at: https://www.cmap.illinois.gov/documents/10180/126764/_Combined_AllCCAs.pdf/

CNT (2019). Equity and Smart Mobility. Available at: https://www.cnt.org/publications/equity-and-smart-mobility

Cohen, J., Cohen, P., West, S.G., & Aiken, L.S. (1983). Applied multiple regression. *Correlation Analysis for the Behavioral Sciences*, *2*.

Coleman, N., Esmalian, A., & Mostafavi, A. (2020). Equitable resilience in infrastructure systems: empirical assessment of disparities in hardship experiences of vulnerable populations during service disruptions. *Natural Hazards Review*, 21(4), p.04020034.

Correa, D., Xie, K., & Ozbay, K. (2017). Exploring the taxi and Uber demand in New York City: An empirical analysis and spatial modeling. In 96th *Annual Meeting of the Transportation Research Board*, Washington, DC.

CTA (2019). Annual Ridership Report: January 2019. Available at: https://www.transitchicago.com/assets/1/6/2018_Annual_Report_-_v3_04.03.2019.pdf

CTA (2020a). Facts at a Glance. Available at: https://www.transitchicago.com/facts/

CTA (2020b). Monthly Ridership Report: December 2019. Available at: https://www.transitchicago.com/assets/1/6/Monthly_Ridership_2019-12.pdf

Dargin, J.S., & Mostafavi, A. (2020). Human-centric infrastructure resilience: Uncovering well-being risk disparity due to infrastructure disruptions in disasters. *PloS One*, 15(6), p.e0234381.

Daus Esq., M.W. (2016). The Expanding Transportation Network Company "Equity Gap": Adverse Impacts on Passengers with Disabilities, Underserved Communities, the Environment & the On-Demand Workforce. *University Transportation Research Center*, 1-73.

Deka, D., & Fei, D. (2019). A comparison of the personal and neighborhood characteristics associated with ridesourcing, transit use, and driving with NHTS data. *Journal of Transport Geography*, 76, 24-33.

Dias, F. F., Lavieri, P. S., Garikapati, V. M., Astroza, S., Pendyala, R. M., & Bhat, C. R. (2017). A behavioral choice model of the use of car-sharing and ride-sourcing services. *Transportation*, 44(6), 1307-1323.

Divvy (2020). Station Map and Data. Available at: https://member.divvybikes.com/map/

DiPietro, G.S., Scott Matthews, H., & Hendrickson, C.T. (2014). Estimating economic and resilience consequences of potential navigation infrastructure failures: A case study of the Monongahela River. *Transportation Research Part A: Policy and Practice*, 69, 142–164

Donovan, B., & Work, D. B. (2017). Empirically quantifying city-scale transportation system resilience to extreme events. *Transp. Res. Part C*, 79, 333–346. https://doi.org/10.1016/j.trc.2017.03.002.

El-Assi, W., Mahmoud, M.S., & Habib, K.N. (2017). Effects of built environment and weather on bike sharing demand: A station level analysis of commercial bike sharing in Toronto. *Transportation*, 44(3), pp. 589-613.

Enders, C.K., & Tofighi, D. (2007) 'Centering Predictor Variables in Cross-sectional Multilevel Models: A New Look at an Old Issue', *Psychological Methods*, 12: 121–38

Estévez-Mauriz, L., Fonseca, J. A., Forgaci, C., & Björling, N. (2017). The livability of spaces: Performance and/or resilience? Reflections on the effects of spatial heterogeneity in transport and energy systems and the implications on urban environmental quality. *International Journal of Sustainable Built Environment*, 6(1), 1-8.

Faturechi, R., & Miller-Hooks, E. (2015). Measuring the performance of transportation infrastructure systems in disasters: A comprehensive review. *Journal of Infrastructure Systems*, 21(1), p.04014025.

Fotouhi, H., Moryadee, S., &Miller-Hooks, E. (2017). Quantifying the resilience of an urban traffic-electric power coupled system. *Reliab. Eng. Syst. Saf.* 163, 79–94. https://doi.org/10.1016/j.ress.2017.01.026.

Ferrell, C., Appleyard, B., & Taecker, M. (2016) A Handbook for Building Livable Transit Corridors: Methods, Metrics and Strategies. (No. TCRP H-45) *Transit Cooperative Research Program*.

Fuller, D., Sahlqvist, S., Cummins, S., & Ogilvie, D. (2012). The impact of public transportation strikes on use of a bicycle share program in London: Interrupted time series design. *Preventive medicine*, 54(1), 74-76.

Gehrke, S.R., Felix, A., & Reardon, T.G., (2019). Substitution of ride-hailing services for more sustainable travel options in the greater Boston region. *Transportation Research Record*, 2673(1), pp.438-446.

Ghaffar, A., Mitra, S., & Hyland, M. (2020). Modeling determinants of ridesourcing usage: A census tract-level analysis of Chicago. *Transportation Research Part C: Emerging Technologies*, *119*, 102769.

Ge, Y., Knittel, C. R., MacKenzie, D., & Zoepf, S. (2016). Racial and gender discrimination in transportation network companies (No. w22776). *National Bureau of Economic Research*.

Gelman, A., & Hill, J. (2007). Data Analysis Using Regression and Multilevel/Hierarchical Models. Cambridge: Cambridge University Press.

Gill, J., & Womack, A. J. (2013). The multilevel model framework. The SAGE Handbook of Multilevel Modeling. London: SAGE Publications Ltd.

Goldstein, H. (2003). Multilevel Statistical Models (3rd ed). London: Edward Arnold.

Google (2019). Chicago, Illinois. Retrieved from: https://www.google.com/maps/place/Chicago,+IL/@41.8339032,-87.8723909,11z/data=!3m1!4b1!4m5!3m4!1s0x880e2c3cd0f4cbed:0xafe0a6ad09c0c000!8m2!3d41.8781136!4d-87.6297982

Graham, M. (2011). Time machines and virtual portals: The spatialities of the digital divide. *Progress in Development Studies*, 11(3), 211-227.

Grossman, A. (May 1, 2020). Resiliency of Transit Agency Partnerships with Private Mobility Services in the Face of COVID-19. *ENO Center for Transportation*. Available at:

https://www.enotrans.org/article/resiliency-of-transit-agency-partnerships-with-private-mobility-services-in-the-face-of-codiv-19/

Gu, Y., Fu, X., Liu, Z., Xu, X., & Chen, A. (2020). Performance of transportation network under perturbations: Reliability, vulnerability, and resilience. *Transportation Research Part E: Logistics and Transportation Review*, 133, p.101809.

Hughes, R., & MacKenzie, D. (2016). Transportation network company wait times in Greater Seattle, and relationship to socioeconomic indicators. *Journal of Transport Geography*, 56, 36-44.

Iseki, H., Liu, C., & Knaap, G. (2018). The determinants of travel demand between rail stations: A direct transit demand model using multilevel analysis for the Washington DC Metrorail system. *Transportation Research Part A: Policy and Practice*, 116, 635-649.

Jin, S. T., Kong, H., & Sui, D. Z. (2019). Uber, Public Transit, and Urban Transportation Equity: A Case Study in New York City. *The Professional Geographer*, 71(2), 315-330.

Jin, S. T., Kong, H., Wu, R., & Sui, D. Z. (2018). Ridesourcing, the sharing economy, and the future of cities. *Cities*, 76, 96-104.

Jones, K., & Duncan, C. (1996). People and places: the multilevel model as a general framework for the quantitative analysis of geographical data. *Spatial analysis: Modelling in a GIS environment*, 79-104.

Julian, M. (2001). The consequences of ignoring multilevel data structures in nonhierarchical covariance modeling. *Structural Equation Modeling, 8*, 325-352

Kashef, M. (2016). Urban livability across disciplinary and professional boundaries. *Frontiers of Architectural Research*, *5*(2), 239-253.

Kaviti, S., Venigalla, M. M., Zhu, S., Lucas, K., & Brodie, S. (2018). Impact of pricing and transit disruptions on bikeshare ridership and revenue. *Transportation*, 1-22.

King, D., & Shalaby, A. (2016). Performance Metrics and Analysis of Transit Network Resilience in Toronto. In: *TRB 95th Annual Meeting Compendium of Papers*. Washington, D.C.

Krapp, A. (2020). Transportation Equity Project: Prioritization Criteria. *University of Illinois at Urbana-Champaign*.

Lavieri, P. S., Dias, F. F., Juri, N. R., Kuhr, J., & Bhat, C. R. (2018). A model of ridesourcing demand generation and distribution. *Transportation Research Record*, 2672(46), 31-40.

Leng, N., & Corman, F. (2020). The role of information availability to passengers in public transport disruptions: An agent-based simulation approach. *Transportation Research Part A: Policy and Practice*, 133, 214-236.

Leu, G., Abbass, H., & Curtis, N. (2010). Resilience of ground transportation networks: A case study on Melbourne. In: *33rd Australasian Transport Research Forum*, Camberra.

Liu, R., Palm, M., Shalaby, A., & Farber, S. (2020). A social equity lens on bus bridging and ride-hailing responses to unplanned subway disruptions. *Journal of Transport Geography*, 88, 102870.

Litman, T. (2007). Evaluating rail transit benefits: A comment. *Transport Policy*, 14(1), pp.94-97.

Lowe, K. (2014). Bypassing equity? Transit investment and regional transportation planning. *Journal of Planning Education and Research*, 34(1), 30-44.

Lu, Y., Zhao, J., Wu, X., & Lo, S.M. (2021). Escaping to nature during a pandemic: A natural experiment in Asian cities during the COVID-19 pandemic with big social media data. *Science of the Total Environment*, 777, p.146092.

Mahmassani, H.S., & Liu, Y.H. (1999). Dynamics of commuting decision behavior under advanced traveler information systems. *Transportation Research Part C: Emerging Technologies*, 7(2-3), pp.91-107.

Marquet, O. (2020). Spatial distribution of ride-hailing trip demand and its association with walkability and neighborhood characteristics. *Cities*, 106, 102926.

Marsden, G., & Docherty, I. (2013). Insights on disruptions as opportunities for transport policy change. *Transportation Research Part A: Policy and Practice*, 51, 46-55.

Mattsson, L. G., & Jenelius, E. (2015). Vulnerability and resilience of transport systems–A discussion of recent research. *Transportation Research Part A: Policy and Practice*, 81, 16-34.

McFarlane, C. (2011). Assemblage and critical urbanism. *City*, *15*(2), 204-224.

Meerow, S., & Newell, J.P., 2019. Urban resilience for whom, what, when, where, and why? *Urban Geography*, 40(3), pp.309-329.

Mojica, C. H. (2008). Examining changes in transit passenger travel behavior through a smart card activity analysis. Doctoral dissertation, MIT.

National Academies of Sciences, Engineering, and Medicine. (2021a). Transportation System Resilience: Research Roadmap and White Papers. Washington, DC: The National Academies Press.https://doi.org/10.17226/26160.

National Academies of Sciences, Engineering, and Medicine. (2021b). Investing in Transportation Resilience: A Framework for Informed Choices. *TRB Special Report*, 350. Washington, DC: The National Academies Press. Available at: https://doi.org/10.17226/26292.

Nguyen-Phuoc, D., Currie, G., De Gruyter, C., & Young, W. (2018a). How do public transport users adjust their travel behaviour if public transport ceases? A qualitative study. *Transportation Research Part F*, 54, 1–14. https://doi.org/10.1016/j.trf.2018.01.009.

Pangbourne, K., Mladenović, M.N., Stead, D., & Milakis, D. (2020). Questioning mobility as a service: Unanticipated implications for society and governance. *Transportation Research Part A: Policy and Practice*, 131, pp.35-49.

Pnevmatikou, A. M., Karlaftis, M. G., & Kepaptsoglou, K. (2015). Metro service disruptions: How do people choose to travel? *Transportation*, 42(6), 933-949.

Pu, W., McCall, N., Seifu, M., Hampton, B., Milone, R., Griffiths, R., & Meese, A. J. (2017). State of transportation in a day without metro in the Washington region (No. 17-00132).

R Core Team (2021). R: A language and environment for statistical computing. R Foundation for Statistical Computing, Vienna, Austria. URL https://www.R-project.org/.

Rahimi, E., Shamshiripour, A., Shabanpour, R., Mohammadian, A., & Auld, J. (2020). Analysis of Transit Users' Response Behavior in Case of Unplanned Service Disruptions. *Transportation Research Record*, 2674(3), 258-271.

Rayle, L., Dai, D., Chan, N., Cervero, R., & Shaheen, S. (2016). Just a better taxi? A survey-based comparison of taxis, transit, and ridesourcing services in San Francisco. *Transport Policy*, 45, 168-178.

Saberi, M., Ghamami, M., Gu, Y., Shojaei, M. H. S., & Fishman, E. (2018). Understanding the impacts of a public transit disruption on bicycle sharing mobility patterns: A case of Tube strike in London. *Journal of Transport Geography*, 66, 154-166.

Sarker, R. I., Kaplan, S., Mailer, M., & Timmermans, H. J. (2019). Applying affective event theory to explain transit users' reactions to service disruptions. *Transportation Research Part A: Policy and Practice*, 130, 593-605.

Serulle, N., Heaslip, K., Brady, B., Louisell, W., & Collura, J. (2011). Resiliency of transportation network of Santo Domingo, Dominican Republic. *Transp. Res. Rec. J. Transp. Res. Board* 2234, 22–30. https://doi.org/10.3141/2234-03.

Shaheen, S., Bell, C., Cohen, A., & Yelchuru, B. (2017). Travel behavior: Shared mobility and transportation equity (No. PL-18-007).

Shamshiripour, A., Rahimi, E., Shabanpour, R., & Mohammadian, A. K. (2020). Dynamics of travelers' modality style in the presence of mobility-on-demand services. *Transportation Research Part C: Emerging Technologies*, 117, 102668.

Shared-Use Mobility Center & National Center for Mobility Management. (2020). "Case Study: Examples of Mobility on Demand". Available at: https://learn.sharedusemobilitycenter.org/casestudy/mod-and-accessibility-policies-and-programs-2020/

Sikder, S. (2019). Who uses ride-hailing services in the United States? *Transportation Research Record*, 2673(12), 40-54.

Snijders, T. A., & Bosker, R. J. (1999). Multilevel analysis: An introduction to basic and advanced multilevel modeling. London: SAGE Publications.

Soria, J., Chen, Y., & Stathopoulos, A. (2020). K-Prototype Segmentation Analysis on Large-scale Ridesourcing Trip Data. *Transportation Research Record*. 0361198120929338.

Soria, J., & Stathopoulos, A. (2021). Investigating Socio-spatial Differences between Solo Ridehailing and Pooled Rides in Diverse Communities, Volume 95, 103148, *Journal of Transport Geography*.

StataCorp. (2019). Stata Statistical Software: Release 16. College Station, TX: StataCorp LLC.

Sun, H., Wu, J., Wu, L., Yan, X., & Gao, Z. (2016). Estimating the influence of common disruptions on urban rail transit networks. *Transportation Research Part A: Policy and Practice*, 94, 62-75.

U.S. Census Bureau. (2017). 2017 American Community Survey 5-Year Estimates, *American FactFinder*. Available at: http://factfinder.census.gov

Van Exel, N. J. A., & Rietveld, P. (2001). Public transport strikes and traveller behaviour. *Transport Policy*, 8(4), 237-246.

Van Exel, N. J. A., & Rietveld, P. (2009). When strike comes to town… anticipated and actual behavioural reactions to a one-day, pre-announced, complete rail strike in the Netherlands. *Transportation Research Part A: Policy and Practice*, 43(5), 526-535.

Walker, B., Carpenter, S., Anderies, J., Abel, N., Cumming, G., Janssen, M., Norberg, J., Peterson, G. D., & Pritchard, R. (2002). Resilience Management in Social-ecological Systems: A Working Hypothesis for a Participatory Approach. *Conserv. Ecol.*, 6, 14.

Walker, B., Holling, C.S., Carpenter, S.R., & Kinzig, A. (2004). Resilience, adaptability and transformability in social – ecological systems. *Ecol. Soc.* 9, 5. https://doi.org/10.1103/PhysRevLett. 95.258101.

Wampold, B. E., & Serlin, R. C. (2000). The consequence of ignoring a nested factor on measures of effect size in analysis of variance. *Psychological Methods, 5*(4), 425–433. https://doi.org/10.1037/1082-989X.5.4.425

Wang, M., & Mu, L. (2018). Spatial disparities of Uber accessibility: An exploratory analysis in Atlanta, USA. Computers, *Environment and Urban Systems*, 67, 169-175.

Wen, H., Xiao, Y., & Zhang, L. (2017). School district, education quality, and housing price: Evidence from a natural experiment in Hangzhou, China. *Cities*, 66, pp.72-80.

Younes, H., Nasri, A., Baiocchi, G., & Zhang, L. (2019). How transit service closures influence bikesharing demand; lessons learned from SafeTrack project in Washington, DC metropolitan area. *Journal of Transport Geography*, 76, 83-92.

Young, M., & Farber, S. (2019). The who, why, and when of Uber and other ride-hailing trips: An examination of a large sample household travel survey. *Transportation Research Part A: Policy and Practice*, 119, pp.383-392.

Yu, H., & Peng, Z. R. (2019). Exploring the spatial variation of ridesourcing demand and its relationship to built environment and socioeconomic factors with the geographically weighted Poisson regression. *Journal of Transport Geography*, 75, 147-163.

Zhang, Y., & Zhang, Y. (2018). Examining the relationship between household vehicle ownership and ridesharing behaviors in the United States. *Sustainability*, 10(8), 2720.

Zhao, F., Chow, L. F., Li, M. T., Ubaka, I., & Gan, A. (2003). Forecasting transit walk accessibility: Regression model alternative to buffer method. *Transportation Research Record*, 1835(1), 34-41.

Zhu, S., Masud, H., Xiong, C., Yang, Z., Pan, Y., & Zhang, L. (2017). Travel Behavior Reactions to Transit Service Disruptions: Study of Metro SafeTrack Projects in Washington, DC. *Transportation Research Record*, 2649(1), 79-88.